\documentclass[aip,jap,reprint]{revtex4-1}

\draft % marks overfull lines with a black rule on the right

\usepackage{graphicx}% Include figure files
\usepackage{dcolumn}% Align table columns on decimal point
\usepackage{bm}% bold math
\usepackage{color}% bold math
\usepackage{float}
\usepackage{hyperref}
\usepackage{mhchem}

\begin{document}

\title{Strong microwave photon-magnon coupling in multiresonant dielectric antennas (Perspective)} 
\author{Ivan S. Maksymov}
\email{ivan.maksymov@rmit.edu.au}
\affiliation{Australian Research Council Centre of Excellence for Nanoscale BioPhotonics, School of Science, RMIT University, Melbourne, Victoria 3001, Australia}
\affiliation{Department of Physics, University of Western Australia, 35 Stirling Highway, Crawley WA 6009, Australia}

\date{\today}

\begin{abstract}

Achieving quantum-level control over electromagnetic waves, magnetisation dynamics, vibrations and heat is invaluable for many practical application and possible by exploiting the strong radiation-matter coupling. Most of the modern strong microwave photon-magnon coupling developments rely on the integration of metal-based microwave resonators with a magnetic material. However, it has recently been realised that all-dielectric resonators made of or containing magneto-insulating materials can operate as a standalone strongly-coupled system characterised by low dissipation losses and strong local microwave field enhancement. Here, after a brief overview of recent developments in the field, I discuss examples of such dielectric resonant systems and demonstrate their ability to operate as multiresonant antennas for light, microwaves, magnons, sound, vibrations and heat. This multiphysics behaviour opens up novel opportunities for the realisation of multiresonant coupling such as, for example, photon-magnon-phonon coupling. I also propose several novel systems in which strong photon-magnon coupling in dielectric antennas and similar structures is expected to extend the capability of existing devices or may provide an entirely new functionality. Examples of such systems include novel magnetofluidic devices, high-power microwave power generators, and hybrid devices exploiting the unique properties of electrical solitons. 

\end{abstract}

\maketitle %\maketitle must follow title, authors, abstract and \pacs

\section{Introduction}

Novel technologies enabling controllable and efficient interactions of electromagnetic radiation with matter are central to achieving the ambitious goal of quantum information processing \cite{Fey82, Sho97} with light, microwaves, magnetisation dynamics, vibrations and heat \cite{Xia13, Lop13, Gor14, Gus14, Bar15, Xia17, Cam18}. These technologies also advance our capability to develop novel biomedical imaging modalities \cite{Dru17}, frequency-tuneable metamaterials \cite{Lap14, Gre14}, and radars \cite{Lan11, Luk16}.

A practicable physical system enabling strong radiation-matter interactions has to be able to exchange information with preserved coherence. Such a system has to operate in a strong coupling regime, where the coupling strength between two or more subsystems is larger than the mean energy loss in both of them \cite{Jay63, Tav68, Wan15}.

A natural way to increase the coupling strength and decrease energy losses is to employ a resonant system \cite{Xia13}. For example, in photonics, strong coupling has been achieved by forcing electric dipoles of semiconductor quantum dots or nitrogen-vacancy centres in diamonds to interact with optical fields of photonic resonators \cite{Xia13}. Strong coupling has also been demonstrated with magnetic dipoles \cite{Ima09}, for example, by using a microwave resonator loaded with a magnetic material. Here, the interaction of magnons -- collective oscillations in the alignment of spins in the magnetic material -- with microwave resonator photons results in strongly hybridised magnon-resonator states also called magnon-polaritons \cite{Xia13, Zha_NatCommun}. 

\begin{figure}[H]
\centerline{
\includegraphics[width=8cm]{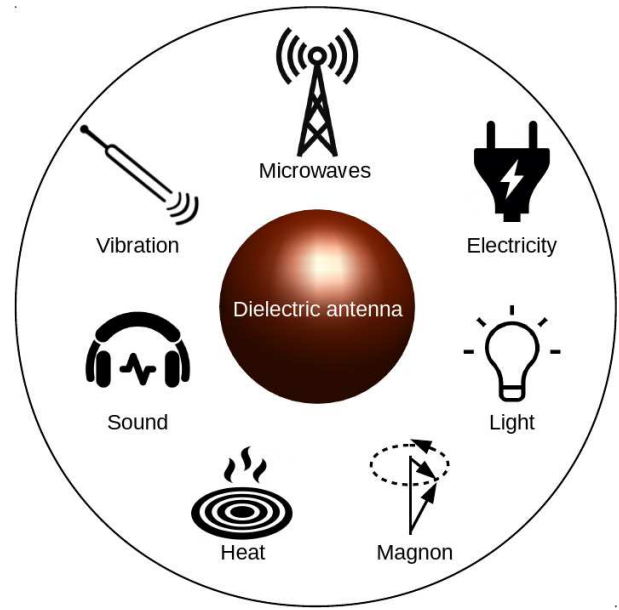}}
\caption{Conceptual illustration of a multiresonant dielectric antenna that mediates the coupling between different entities such as magnons, electricity, microwaves and light, sound and vibrations, and heat. In this role, the same dielectric antenna supports electromagnetic (microwave and optical frequencies), ferromagnetic and acoustic resonances, as well as can be controlled by electrical currents (e.g. via conductive electrical contacts) and electrical fields. Although an yttrium iron garnet (YIG) sphere is pictured because of its obvious role as a resonant dielectric antenna (Sec.~\ref{YIG sphere as antenna}), other dielectric structures are also conceivable as discussed in the main text.}
\label{fig:fig0}
\end{figure}

Very often, strong microwave photon-magnon coupling can be achieved with microwave resonators such as a metal box \cite{Tab14, Zha14}, in which microwaves are reflected and trapped between the metal walls. Such resonators have high quality factors and are relatively simple to construct and theoretically analyse, which makes them indispensable for microwave devices \cite{Poz98}.

However, the walls of a box resonator are a significant obstacle for light, sound, vibrations and heat. Although innovative techniques enabling the coupling of light, sound and heat to the magnetic material located inside a metal resonator have been demonstrated\cite{Zha16_2, His16}, the search remains open for alternative resonant structures. 

In this perspective, I present an emergent approach that utilises all-dielectric resonant structures, also generally called dielectric antennas \cite{Kra13, Key16}, to control strongly hybridised magnon-polaritons with electrical currents, light, sound, vibrations and heat (Fig.~\ref{fig:fig0}). In this approach, a dielectric antenna simultaneously operates as a resonator for magnons, microwaves, light, sound and vibrations, as well as it can be controlled with heat, electric fields and currents. This represents a shift in the existing design of strongly-coupled systems and ushers new classes of devices based on a multiresonant strong coupling between, for example, photons, magnons and phonons \cite{Zha16_2, His16}. Throughout the paper, I discuss the recent developments in this very active field, suggest future research directions and propose several novel systems in which the application of dielectric antennas and similar structures is likely to extend the capability of existing devices or provide an entirely new functionality.

\begin{figure}[t]
\centerline{
\includegraphics[width=8cm]{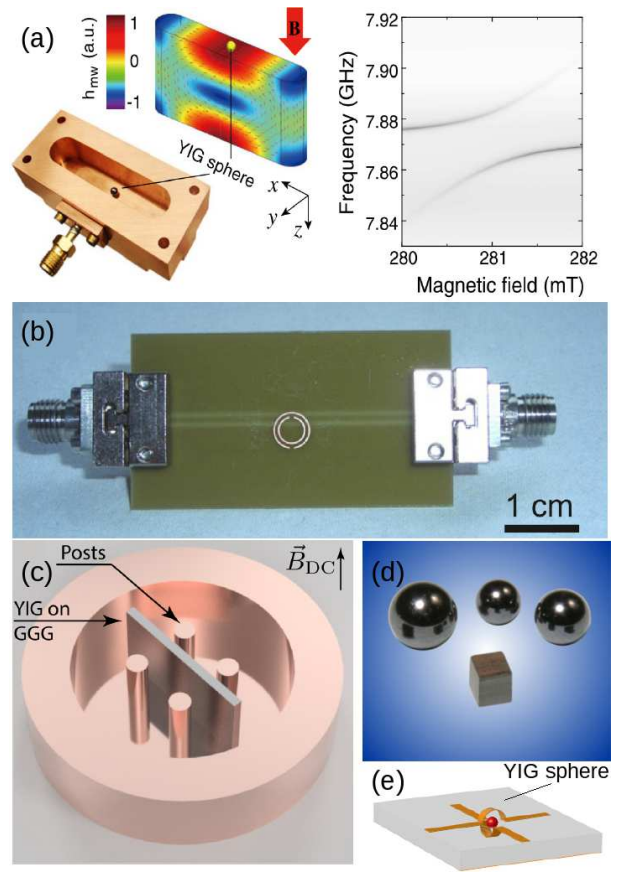}}
\caption{(a) Left: Microwave magnetic field distribution in a microwave metal resonator tuned on the TE$_{\rm{101}}$ mode. Image showing one half of the microwave resonator with an YIG sphere inside, which is located in the area where the microwave magnetic field amplitude reaches its maximum. Right: Measured anti-crossing of the microwave photon mode and the FMR mode of the YIG sphere as a function of the applied magnetic field, which indicates a strong microwave photon-magnon coupling. (b) Microwave split-ring resonator (SRR) printed on top of a coplanar microwave line. The SRR is to be loaded with an YIG film to achieve strong microwave photon-magnon coupling. (c) Four post re-entrant metal resonator loaded with an YIG film. (d) Commercial YIG-sphere resonators (courtesy of Ferrite Domen Co.) Spheres with the diameter from $0.25$~mm to $5$~mm are available. YIG spheres are produced by slowly shaping YIG cubes into a YIG sphere. (e) Schematic of an YIG-sphere resonator coupled to a microstrip circuit. Reproduced with permission from Refs.~\onlinecite{Zha14, Gre14, Gor18}. The image in (e) is from Wikimedia Commons.}
\label{fig:fig1}
\end{figure}

\section{Strong coupling with microwave metal resonators}

From its very birth, the field of strong microwave photon-magnon coupling has relied on metal resonators loaded with a magnetic material. Therefore, in this section I briefly overview the recent progress achieved with metal resonator structures.

The interaction of a single spin with a microwave photon is intrinsically weak \cite{Xia13}. However, collective enhancement, which scales as the square root of the total number of spins, results in a considerable increase in the coupling strength \cite{Ima09}. The spin systems with the largest volumetric density of spins are ferro/ferrimagnets, such as yttrium iron garnet (YIG) \cite{Che93}. In contrast to paramagnetic spins \cite{Xia13}, in ferrimagnetic materials, exchange and dipole-dipole interactions between the spins result in a collective motion of spins, which leads to the ferromagnetic resonance (FMR) \cite{YIG_review, Mak15}. In this picture, magnons represent the quanta of the FMR and their frequency depends on the static magnetic field applied to the ferrimagnetic material.

The collective behaviour of spins, their large density, and extremely small magnetic losses in ferrimagnets \cite{Che93} result in a strong response of these materials to an external microwave field. As a result, even small volumes of a bulk ferromagnetic material allow achieving high-cooperativity quantum electrodynamics (QED) in hybrid resonant systems \cite{Tab14, Zha14, Bho14, Gor14, Zha15, Hai15, Kos16, Lam16, Zha17, Bho17, Mor17, Cas17, Rao17, Ram18, Wan18, Abd18}. Very often, this is achieved by using a microwave metal resonator loaded with a single-crystal YIG sphere \cite{Soy10, Tab14, Zha14, Gor14, Zha15, Bou16} [Fig.~\ref{fig:fig1}(a), left]. YIG spheres can also be excited with microwave stripline resonators \cite{Rao17, Cas17, Mor17} and coaxial cable resonators \cite{Hai15, Lam16}.

A strong anti-crossing of the microwave resonator mode and the FMR mode of the YIG sphere as a function of the applied magnetic field indicates a strong coupling between the microwave photon and the magnon modes [Fig.~\ref{fig:fig1}(a), right]. This effect can be observed at room temperatures. A rectangular YIG crystal was also used to demonstrate the same result \cite{Hue13, Yao15}. However, whereas with YIG spheres one avoids the effects of an inhomogeneous demagnetisation field, such effects are present in rectangular resonators, which leads to the unwanted coupling of many magnetostatic modes to the resonator modes.

Unfortunately, the YIG-sphere technology is essentially $3$D and hence it may be incompatible with planar platforms. Therefore, a microwave stripline resonator or a split-ring resonator [Fig.~\ref{fig:fig1}(b)] loaded with an YIG film have been proposed \cite{Xia13, Ste13, Gre14, Bho14, Bho17, Cas17}. YIG films have also been employed inside metal microwave resonators \cite{Cao15, Zha16, Dod17, Gor18} [Fig.~\ref{fig:fig1}(c)], where they may have additional technological advantages over YIG spheres \cite{Gor18}.    

\section{Strong coupling with dielectric antennas}

\subsection{Dielectric microwave antennas}

In many practical applications, dielectric microwave resonators are a viable alternative to metal ones because of lower losses at high microwave frequencies \cite{Oka62}. For example, dielectric resonators are well-known in the areas of electron spin resonance \cite{Har69}, ferromagnetic resonance \cite{Ser07, Byc13}, magnetically-tuneable metamaterials \cite{Bi14, Sun15} and similar structures \cite{Nik14, Nik17, Abd18}. (However, in the cited papers, strong coupling was not achieved as discussed elsewhere \cite{Mak15_slot, Ram15}.)

Another advantage of the dielectric resonators over the metal ones, which also becomes more pronounced at high microwave frequencies, is the absence of eddy currents induced by microwave magnetic fields in the metal parts \cite{Bai13, Mak13, Mak14}. For example, the eddy-current shielding of microwave magnetic fields have been predicted \cite{Mak13} to have an adverse effect on the operation of YIG-platinum bi-layer structures, which have recently been proposed to detect and control magnon-polaritons with electrical currents and heat \cite{Cao15, Cas17}. Therefore, the use of dielectric resonators in systems with strong coupling would be advantageous.

In the field of radio-frequency antennas, dielectric resonator structures are also known as dielectric microwave antennas \cite{Key16}. Similar to microwave dielectric resonators, an advantage of dielectric antennas over conventional metal antennas is that they do not have metal parts that often dissipate energy at high microwave frequencies \cite{Key16}. Dielectric antennas made of a high dielectric permittivity ($\epsilon$) materials may also be more compact than metal antennas, because the size of the dielectric antennas scales as $\lambda_{\rm{0}}/\sqrt{\epsilon}$ being $\lambda_{\rm{0}}$ the free-space wavelength of the incident wave.

\subsection{YIG spheres as microwave dielectric antennas} \label{YIG sphere as antenna}

A dielectric antenna may consist of a single sphere made of a material with a large $\epsilon$ \cite{Fil12, Kuz13, Kra13}. Several dielectric spheres of different radii may be combined to build a more complex antenna such as, for example, a directive Yagi-Uda antenna \cite{Kra14}. Here, spheres made of low-loss ceramics with $\epsilon=16$ at $9-12$~GHz may be employed \cite{Kra14}.

Significantly, YIG spheres [Fig.~\ref{fig:fig1}(d, e)] can operate as an independent microwave resonator antenna because at microwave frequencies $\epsilon_{\rm{YIG}}\approx15$ \cite{Che93}. Therefore, by using a single YIG sphere, one may achieve the strong photon-magnon coupling without an external resonator by confining the photon and magnon modes in the volume of the sphere \cite{Ram15, Bou16}. In this case, the diameter of the YIG sphere antenna should be so chosen that a microwave Mie resonances inside the sphere is excited \cite{Ram15}.

Not only single-sphere YIG antennas may be used to achieve strong coupling. For example, intriguing effects such as long-range strong coupling of magnons in spatially separated YIG spheres has been observed \cite{Lam16, Ram18}. Thus far, this interaction has been mediated by a microwave resonator. However, one may employ YIG spheres without an external resonator by tuning the spheres to operate as a dielectric antenna \cite{Kra13}.     

In dielectric YIG antennas, the frequency of the mode anti-crossing due to the strong coupling may be increased by decreasing the sphere radius. However, due to a relatively low saturation magnetisation of YIG, this would also require larger static magnetic fields. Lower operating fields will be attainable by using spinel or hexaferrite materials \cite{Har09} instead of YIG. Such materials are characterised by a large saturation magnetisation of $\sim3000-5000$~G, low magnetic losses comparable to those of YIG, and $\epsilon\approx12$. 

\begin{figure}[t]
\centerline{
\includegraphics[width=8cm]{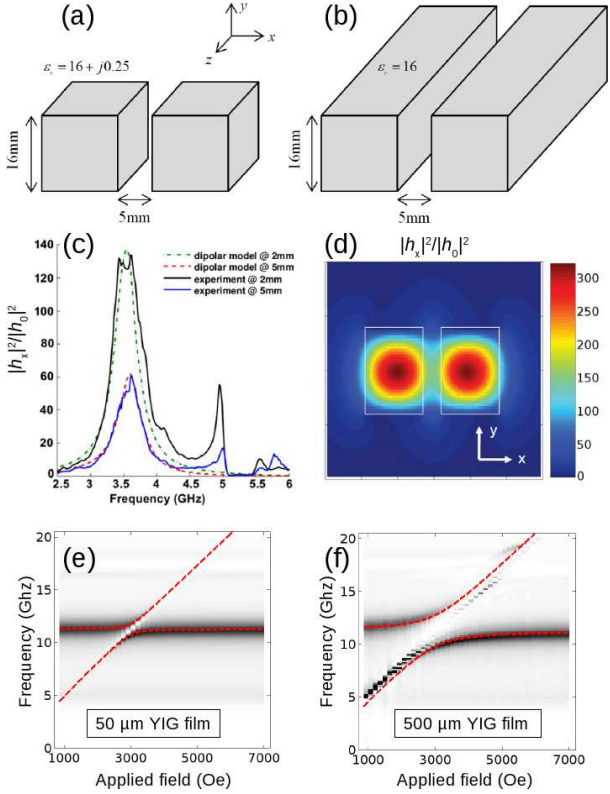}}
\caption{(a) Schematic of the dielectric antenna proposed by G. Boudarham \textit{et al.} \cite{Bou14} to enhance the microwave magnetic field in the gap between the two dielectric cubes. (b) Schematic of the waveguide antenna proposed to enhance the in the gap between the two dielectric bars \cite{Nam15}. (c) Experimental (full lines) and theoretical (dashed lines) normalised magnetic field intensities in the gap of the antenna for the gap width of $5$~mm and $2$~mm. The magnetic field component of the excitation wave is polarised along the \textit{x}-axis. Reproduced with permission from Ref.~\onlinecite{Bou14}. (d) Magnetic field enhancement in the waveguide antenna simulated using an FDTD method \cite{Mak13}. Straight white lines denote the contours of the waveguide cross-section. (e, f) The anti-crossing effect (simulation) in the waveguide antenna consisting of two finite-length dielectric bars separated by an YIG film. The red dashed curves denote the best fit with a model of two coupled resonators.}
\label{fig:fig3}
\end{figure}

\subsection{YIG dielectric antennas and microwave magnetic field enhancement} 

In systems enabling a strong light-matter coupling through electric near-field enhancement using electric dipoles, the medium hosting the dipoles (e.g. a quantum dot or a nanodiamond containing nitrogen-vacancy centres) has to be located in the area where the \textit{electric} field in the resonator reaches its maximum \cite{Bab10, Mak10_1}. When magnetic dipoles are exploited \cite{Ima09}, the magnetic medium has also to be placed in the area where the \textit{magnetic} field reaches its maximum [Fig.~\ref{fig:fig1}(a), left]. In both cases, the collocation with the respective field maxima allows increasing the coupling strength \cite{Kra13}.

Consequently, in systems exploiting quantum dots and nitrogen-vacancy centres, a local increase in the electric field intensity was achieved by judiciously engineering nanophotonic resonators \cite{Bab10, Mak10_1} and optical nanoantennas \cite{Kra13}. By analogy, a local magnetic field enhancement would be favourable for a stronger hybridisation of microwave photons and magnons. However, in both visible and microwave spectral ranges, it is more challenging to enhance the magnetic field than the electric one \cite{Kra13, Kuz13, Bi14, Sun15, Kap17}.

To achieve a strong microwave magnetic field enhancement, G. Boudarham \textit{et al.} \cite{Bou14} demonstrated a dielectric gap antenna that consists of two dielectric cubes separated by a small, sub-wavelength air gap [Fig.~\ref{fig:fig3}(a)]. Measurements were conducted in the GHz spectral range where $\epsilon$ of the cubes is $16+i0.25$. A strong, $\sim140$-fold magnetic field intensity enhancement [Fig.~\ref{fig:fig3}(c)] was demonstrated in the gap between the cubes, and a stronger enhancement was observed for smaller gap widths.

A similar antenna structure consisting of square cross-section bars made of a high dielectric permittivity material separated by an air gap [Fig.~\ref{fig:fig3}(b)] was proposed by our research group \cite{Nam15}. Being inspired by the dielectric slot waveguides that provide a local \textit{electric} field enhancement in a broad spectral range \cite{Alm04}, the proposed waveguide antenna produces a strong magnetic field enhancement within the gap [Fig.~\ref{fig:fig3}(d)]. However, in contrast to the gap antenna consisting of the two cubes [Fig.~\ref{fig:fig3}(a)], the enhancement effect in the waveguide antenna may be broadband provided that the length of the bars is large \cite{Nam15}. The theoretically predicted magnetic field intensity enhancement in the gap between the two infinitely-long bars is $\sim130$ [Fig.~\ref{fig:fig3}(d)], which is close to the enhancement in Fig.~\ref{fig:fig3}(c). When the length of the bars is finite, the structure operates similar to the antenna consisting of the two cubes, but it also supports higher-order modes at high microwave frequencies. 

An YIG film can be placed in the gap of the waveguide antenna to achieve a strong hybridisation of magnons with microwave resonance modes \cite{Mak15_slot}. In this case, the thickness of the gap is equal to the film thickness (e.g. $50$~$\mu$m). This deep subwavelength, with respect to microwave wavelengths, feature enables a strong, $\sim350$-fold magnetic field intensity enhancement inside the YIG film \cite{Mak15_slot}. Together with a strong spatial overlap between the photon and magnon modes, this enhancement greatly facilitates the anti-crossing between the microwave photon and FMR magnon modes [Figs.~\ref{fig:fig3}(e, f)].

The gray-scale frequency-applied external magnetic field $H$ maps in Figs.~\ref{fig:fig3}(e, f) allow extracting the coupling strength $\Delta$ (measured in frequency units) by using an abstract model of two coupled resonances \cite{Gor14}. From the best fits [the dashed curves in Figs.~\ref{fig:fig3}(e, f)], for the $50$~$\mu$m-thick YIG film placed in the antenna gap one obtains $\Delta = 500$~MHz or $\Delta/f_{\rm{1}}^{\rm{0}}=4.5$\%, where $f_{\rm{1}}^{\rm{0}}$ is the resonance frequency of the electromagnetic fundamental mode of the whole dielectric-YIG-dielectric structure at $H=0$, which is a signature of the strong coupling regime \cite{Wan15}. For the $500$~$\mu$m-thick YIG film, $\Delta = 1500$~MHz or $\Delta/f_{\rm{1}}^{\rm{0}}=13.5$\%, which indicates that the the coupling strength becomes comparable to the resonance frequency of the dielectric antenna and the coupling regimes is ultrastrong \cite{Wan15}. Also note minor anti-crossing effects at the higher frequencies ($\sim16$~GHz and $\sim18$~GHz) corresponding to the higher-order electromagnetic resonance modes of the dielectric antenna structure.

To conclude this section, it is worth noting that the quality factor of dielectric antennas may be not high enough for certain applications. An increase can be achieved by exploiting whispering gallery modes of dielectric spheres and discs or by using novel resonance regimes \cite{Ryb17}.  

\section{Multiresonant dielectric antennas}

This section discusses the concept of a multiresonant dielectric antenna that mediates the coupling between different entities (Fig.~\ref{fig:fig0}): microwaves, magnons, light, electricity (used as a general term defining electrical techniques \cite{Skl16, Ota17}), sound, vibrations, and heat. This section also describes the emergent applications of such antennas.

\subsection{Light}

YIG, magneto-insulating bismuth iron garnet (BIG) and other ferrite materials \cite{Har09} are mostly transparent to infrared and visible light \cite{Stancil, Mak15_1, Fri18}. Therefore, the interaction of light with magnons inside the same dielectric antenna volume has enabled the development of hybrid devices such as microwave spectrum analysers, optical frequency shifters, tunable optical filters, and optical beam deflectors \cite{Stancil}.

Significantly, this also enables intriguing interactions of optical, microwave and magnon modes inside the same dielectric antenna, which allows improving the efficiency of the microwave-to-optical photon conversion for quantum information \cite{Dev13, Tab15, His16, Sha17} and optomagnonics \cite{Zha16_2, Osa16, Liu16, Kus16} applications. 

In many experimental scenarios, where an YIG sphere is used together with a metal microwave cavity to achieve a strong photon-magnon hybridisation [Fig.~\ref{fig:fig1}(a)], special measures have to be taken to modify the walls of the metal cavity to couple light to the YIG sphere \cite{His16}. Alternatively, a specially designed microwave antenna has to be used to simplify the access of light \cite{Zha16_2}. These obstacles are lifted when the YIG sphere operates as an independent resonant structure \cite{Osa16, Zha16_2, Sha17}.

A novel solution to the problem of the magnon-to-light conversion using travelling magnons in the form of magnetostatic spin waves in YIG films \cite{Mak15} has recently been proposed \cite{Kos18}. Because of their strong resonance interaction with the matter, travelling magnons propagate much slower than a normal electromagnetic wave, which can be utilises to achieve a considerable magneto-optical modulation based on the Faraday effect \cite{Mak15_1}. Furthermore, such a microwave-to-optical photon converter is an intrinsically planar device that, unlike YIG spheres, is a promising candidate for applications in integrated optomagnonic chips. The gap dielectric antenna (Fig.~\ref{fig:fig3}) could be another candidate for this application, and it would also enable a local enhancement of both the microwave magnetic field \cite{Nam15} and the optical field \cite{Alm04}.

The optical field enhancement will also be important when the magnon modes are excited directly with light \cite{Che17, Sav17}. In this case, the ultrafast optical field excites the spins in the magneto-insulating material via the inverse Faraday effect \cite{Mak15_1}. The excited spins start to precess along the total magnetic field with a resonance at typical GHz-range FMR mode frequencies. 

\subsection{Electrical methods and heat}

Several research groups have demonstrated electrical methods to detect magnons coupled with microwave photons by placing a hybrid YIG-platinum (Pt) bilayer into a microwave cavity \cite{Bai15, Mai16, Bai16}. New features not observed in any previous spin-pumping experiment but predicted theoretically \cite{Cao15} were revealed in those works. 

V. Castel \textit{et al.} \cite{Cas17} demonstrated thermal control of the magnon-photon coupling at room temperature in a resonant notch filter loaded with a hybrid YIG-Pt system. The thermal control of the photon-magnon hybridisation at the resonant condition has been realised by current-induced Joule heating in the Pt film that results in a temperature gradient. Although this planar configuration has a low quality factor, it allows avoiding the adverse impact of the YIG-Pt stack: in a high-quality-factor resonator the insertion of a hybrid stack including a good electrical conductor (Pt) results in significant losses \cite{Cao15}.

Despite this improvement, the high conductivity of the Pt layer may continue to adversely affect the magnetisation dynamics in the YIG layer due to the eddy current shielding effect \cite{Mak13}. This effect can be mitigated by introducing $2$D micro- and nano-patterning of the Pt layer \cite{Gre14, Zho15}.

The nanopatterning of the conductive layer opens up an additional opportunity for the control of the strong coupling with Joule heat. Metal nanostructures are known to support plasmon resonances -- collective oscillations of the electron charge around metal nanoparticles in resonance with the frequency of the incident light \cite{Kra13}. All metals, including gold and silver that are the key constitutive material for modern plasmonics, considerably absorb light and convert its energy into heat. Platinum is often considered to be a poor plasmonic materials due to its high absorption of light \cite{Kra13}. However, this property turns to be useful for the control of the photon-magnon coupling with heat. The heat produced by light in the nanopatterned Pt layer is estimated to be comparable with that produced in nickel nanostructures (similar to Pt nickel is another poor plasmonic material \cite{Mak16_1}), in which temperatures $\sim 350$~$^{\rm{o}}$C were observed \cite{Kat18}.    

Another electrical method that may be used to control the photon-magnon coupling consists in exploiting the dual electric and magnetic tunability of the physical properties of multiferroic materials \cite{Nik14, Nik17}. A system realising this control mechanism may consists of a ferrite film hybridised with a ferroelectric film. There, a spin-electromagnetic wave is formed as a result of hybridisation of the spin wave in the ferrite film with the incident electromagnetic wave. This hybridisation is electrically and magnetically tuneable \cite{Nik14}.

\subsection{Sound and vibrations} \label{sound and vibrations}

Magnetostriction is a property of ferro- and ferrimagnetic materials to change their shape and dimensions during the magnetisation process caused by the applied magnetic field. The magnetostrictive force provides an alternative mechanism to control the photon-magnon coupling. For example, in YIG structures such as spheres, the varying magnetisation induced inside the sphere causes deformations of its spherical geometry (and vice versa), thereby affecting the photon-magnon coupling \cite{Zha16_1}. In other words, an YIG sphere may simultaneously serve as a resonator for magnons, microwaves and sound (phonons). Some magnetostrictive materials are also transparent for infrared and visible light \cite{Stancil, Suk17}, which opens up opportunities to control hybridised microwave photon-magnon-phonon excitations with light \cite{Mak15_1}.

Whereas the magnetostrictive interaction is associated with strain and stress, single-crystal YIG and spinels are very rigid (the Young's modulus of $E \approx 200$~GPa; for comparison one of the hardest materials -- diamond has $E=1220$~GPa) \cite{Suk17}. Similar to metal nanoantennas that have a slightly smaller $E$ \cite{Mak16}, considerable deformation of an YIG dielectric antenna due to acoustic pressure or other mechanical force is challenging. Consequently, novel approaches are required to achieve larger deformations and these will be described in Sec.~\ref{magnetic fluids}.

\section{Novel hybrid systems and applications}

This section suggests several novel research directions and hybrid resonant systems that employ magneto-insulating materials and may exploit the strong photon-magnon interaction to achieve new functionality. 

\subsection{Electrical solitons}

Solitons are a special class of pulse-shaped waves that propagate in nonlinear dispersive media without changing their shape  \cite{Rem03, Ric11}. Nature offers a variety of soliton examples, including solitary water waves, solitons in optical fibres and spin-wave solitons \cite{Rem03, Ric11}.

\begin{figure}[t]
\centerline{
\includegraphics[width=8cm]{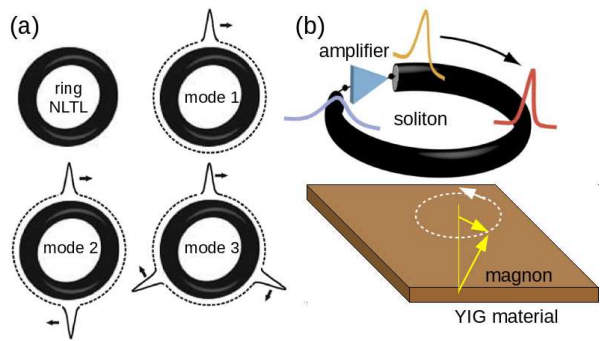}}
\caption{ (a) Ring nonlinear transmission line (NLTL) and electrical soliton modes propagating in it.  (b) Soliton oscillator loaded with an YIG plate. A strong coupling regime would be achieved due to the interaction between the electrical solitons in the NLTL and magnons in the YIG plate. Reproduced with permission from Ref.~\onlinecite{Ric07}.}
\label{fig:fig5}
\end{figure}

Electrical solitons propagate in nonlinear transmission lines (NLTLs) that may be constructed by periodically loading a linear transmission line with varactors \cite{Rem03}. Electrical solitons in two-port NLTLs have been known for several decades \cite{Rem03}. However, electrical solitons in one-port ring NLTLs [Fig.~\ref{fig:fig5}(a)], which self-generate a periodic stable train of electrical soliton pulses, have been demonstrated only recently \cite{Ric07, Ric11}. A ring NLTL supports cnoidal soliton modes [Fig.~\ref{fig:fig5}(a)]. When this NLTL is broken and an amplifier is inserted [Fig.~\ref{fig:fig5}(b)], the initial start-up from noise can compensate for losses in the system \cite{Ric11}, which is commonly done in LC or standing wave oscillators. This leads to the self-generation of self-sustainable soliton modes.

I propose to load a ring NLTL supporting GHz-range repetition rate electrical solitons with an YIG film [Fig.~\ref{fig:fig5}(b)]. In such a hybrid NLTL-YIG multiresonant system, the strong coupling would be manifested as a strong anti-crossing between the soliton modes and the magnon modes in the YIG structure. 

The ring NLTL supporting electrical solitons [Fig.~\ref{fig:fig5}(b)] visually resembles a split-ring resonator [Fig.~\ref{fig:fig1}(b)]. Indeed, a split-ring resonator can be converted into a ring NLTL by adding a nonlinear element (see Fig.~$1$ in Ref.~\onlinecite{Lap14}) and, in principle, it may support solitons \cite{Lap14}. However, the use of an amplifier is a significant step forward from earlier attempts to demonstrate stable solitons in split-ring resonator metamaterials, because in metamaterials solitons did not exist in a pure form but instead soliton-like pulses or dissipative solitons were observed due to nonlinearity \cite{Lap14}.

\subsection{High-power microwave frequency generation}

Apart from important applications in quantum information systems \cite{Gor14}, strong microwave photon-magnon interactions have also been central in the development of low-loss frequency-tuneable metamaterials \cite{Ste13}. In some magnetic metamaterials, the frequency tuning originates from the anti-crossing between the photon and magnon modes [see, e.g., Fig.~\ref{fig:fig1}(a, right)] and its dependence on the external static magnetic field \cite{Ste13, Bho14}. This control mechanism is especially attractive because it is contactless and allows preserving the structural integrity of the device while its characteristics are being tuned. This section demonstrates that this control mechanism may also be useful for the emergent research direction of frequency-tuneable high-power microwave sources.

\begin{figure}[t]
\centerline{
\includegraphics[width=8cm]{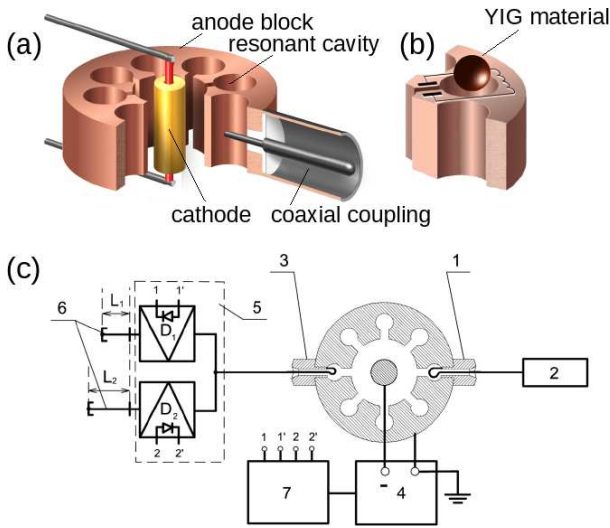}}
\caption{(a) Schematic cross-section of a resonant cavity magnetron. (b) Single resonant cavity in the anode block of the magnetron and its equivalent resonant LC circuit. Ferrite material (schematically shown as a sphere but other geometries are also possible \cite{Neu14}) can be integrated with the cavity). The image of the anode block is adapted from Ref.~\onlinecite{Wolff}. (c) Frequency-tuning scheme of a magnetron with one active power output (1) and one reactive power output (2). A negative polarity voltage from the power supply (4) is applied to the cathode and the power generated by the magnetron is supplied via the active output (1) to an external impedance-matched loading (2). The power from the reactive output (3) is coupled to the switch (5) and two short-circuited waveguides with the length $L_{\rm{1,2}} \leq \lambda_{\rm{WG}}$ being $\lambda_{\rm{WG}}$ the microwave wavelength in the waveguide. The switch diodes are driven by the circuit (7) that is synchronised with the power supply (4). A conceptually similar frequency-tuning scheme may be realised by loading one of the magnetron cavities with a ferrite [as schematically shown in panel (b)], which introduces an extra complex impedance controllable with the an external static magnetic field. Reproduced with permission from Ref.~\onlinecite{Chu16}. }
\label{fig:fig4}
\end{figure}

Frequency-tuneable high-power ($10^{8}-10^{9}$~W) microwave sources are an area of active development \cite{Rom12, Ros10, Hof16, Ahn16, Uma17}. One class of such devices is comprised of NLTLs made of magnetically nonlinear materials such as, for example, ferrites. The NLTL is used to convert a portion of a generally Gaussian or square pulse-like input signal into microwave-frequency energy. When the ferrite is saturated by an external bias field, the propagation of the voltage pulse front gives rise to a shock wave that initiates the magnetic precession in the ferrite and excites the FMR mode. This in turn leads to the modulation of the voltage pulse envelope in the GHz-frequency range.

The NLTL-based high-power microwave devices have several advantages over the conventional vacuum devices \cite{Col11, Ahn15, Gus17}. For example, the NLTLs require no vacuum systems and avoid X-ray production due to high-energy electrons. The NLTLs are also more mechanically rugged that typical vacuum devices.

However, even though electronic biasing of the NLTL's magnetic material may enable some frequency agility that is not readily obtained in many high-power vacuum oscillators \cite{Col11}, in general the ultra-high output energy complicates spectral tuning \cite{Mak10, Hof16}. Another challenge is that the overall efficiency of such sources is also relatively low ($\sim10$\%). In fact, the NLTL output energy consists of a quasi-DC component and a microwave signal. The DC component represents the remainder of the input signal that was not converted into microwave radiation or lost due to attenuation in the line \cite{Hof16}.

The strong photon-magnon coupling and mode anti-crossing effects may be used to control the nonlinear-magnetic interactions in ferrite-based NLTLs, thereby opening up novel opportunities for frequency tuning and efficiency improving. It is noteworthy that magnon-polariton bistability effects have previously been demonstrated in a nonlinear magnonic system consisting of cavity photons strongly interacting with magnon modes of a spherical YIG antenna \cite{Wan18}. The bistable behaviour resulted in a sharp frequency tuning of the magnon polaritons. The same effect would lead to frequency tuning in ferrite-based NLTL high-power microwave generators. Another promising control method would be the use of electric and magnetic fields \cite{Nik14, Nik17} leading to the excitation of spin-electromagnetic waves and frequency tuning.

The strong coupling may also be used to improve cavity magnetron microwave generators. The cavity magnetron is a high-powered vacuum tube that generates microwaves by exploiting the interaction of a stream of electrons with a magnetic field in a series of interconnected resonant metal cavities [Fig.~\ref{fig:fig4}(a)] \cite{magnetron_book, Ben16}.

Magnetrons are the lowest-cost microwave sources in terms of dollars per kW of the generated power, and they have the highest efficiency (typically $>85$\%). However, the frequency and phase stability of magnetrons are not adequate when they are employed as power sources for radar and accelerator applications. The ability to tune the frequency of the generated microwave power is also valuable for radar and electronic warfare applications \cite{Chu16, Chu16_1}. Furthermore, in the case of microwave oven magnetrons, high noise levels at frequencies near the operating frequency $2.45$~GHz presents significant potential for interference with modern wireless telecommunication technologies, which motivates research efforts aimed at developing new frequency-control techniques \cite{Nec03}.

Consequently, novel variable-frequency cavity techniques have been developed to address the problems of phase and frequency locking of magnetrons \cite{Neu10, Neu14, Vya16}. Very often, these techniques use ferrite materials, such as YIG, located inside the resonant cavity of the magnetron [Fig.~\ref{fig:fig4}(b)]. A variable external magnetic field that is orthogonal to the microwave magnetic field of the magnetron varies the permeability of the ferrite, which allows controlling the phase and frequency.  

The strong interaction of microwaves photons and magnons may extend the functionality of this phase and frequency locking scheme. The operating principle of the magnetron requires both static and dynamic magnetic fields, which therefore may be used to excite magnon modes in the ferrite material. The cavities of the magnetron exhibit a resonance analogous to an LC circuit and therefore their loading with the ferrite material is conceptually no different from the hybrid resonant systems such the re-entrant resonator [Fig.~\ref{fig:fig1}(c)]. However, the high microwave power generated by the magnetron will additionally result in nonlinear magnetic effects.

The mode anti-crossing effects may also be used to complement some of the existing frequency-tuning techniques. For example, to tune the output frequency, the anode block of the magnetron is modified by adding an additional microwave energy output Fig.~\ref{fig:fig4}(c). In this scheme, the tuning originates from an electrically-controlled reactive loading of the second output, which is constructively built as an external electric circuit \cite{Chu16, Chu16_1}. As schematically proposed in Fig.~\ref{fig:fig4}(b), by loading one of the anode block resonators with a ferrite material one effectively introduces an extra complex impedance into the resonator [see Fig.~$8$(b) in Ref.~\onlinecite{Bho14} for details]. This extra impedance originates from the magnon modes and its value can be tuned in a wide frequency range by tailoring the photon-magnon coupling with the external magnetic field. This allows reducing constructive changes in the anode block of the magnetron and decreases dependence on external electrical circuits, which would be especially beneficial for high-power magnetron applications.   

To conclude this subsection, it is worth commenting on the suitability of magnetic materials for the proposed application in high-power devices. In general, the power-handling capacity of common ferrites is $\sim 1-2$~kW of CW power \cite{Rizzi}. However, as shown in the papers cited in this subsection, in a nanosecond pulse regime, YIG-based and especially NiZn-based NLTLs can handle much higher powers. The devices utilising NLTLs usually generate damped oscillations with $\sim 5-10$ oscillation periods. Here, the losses are either due to the natural damping of the magnetisation precession \cite{Stancil} or due to the energy dissipation in the waveguide-ferrite system. Thus, the main power handling limitation originates from the breakdown electric field amplitude at the insulator-ferrite interface, but thermal effects become significant mostly at high repetition rates.

\subsection{Strong coupling in magnetofluidic systems} \label{magnetic fluids}

As shown in Sec.~\ref{sound and vibrations}, hybridised microwave photon-magnon-phonon states may be excited in an YIG antenna that supports both electromagnetic and magnon resonances and also can be deformed by an external mechanical force such as sound or vibration. However, in practice, it is challenging to achieve considerable deformations of a solid medium \cite{Mak16}.

In the adjacent area of opto-mechanics, to solve the same problem it was proposed to employ liquid droplets and gas bubbles in liquids \cite{Maa16, Mak16_2, Mak17}. In general, liquid droplets and bubbles oscillate as a function of the external driving force \cite{Ash11}. These effects result in large (and impossible with solid-state materials) changes in the propagation properties of light incident on the medium containing droplets or bubbles \cite{Maa16, Mak16_2, Mak17}.

\begin{figure}[t]
\centerline{
\includegraphics[width=6cm]{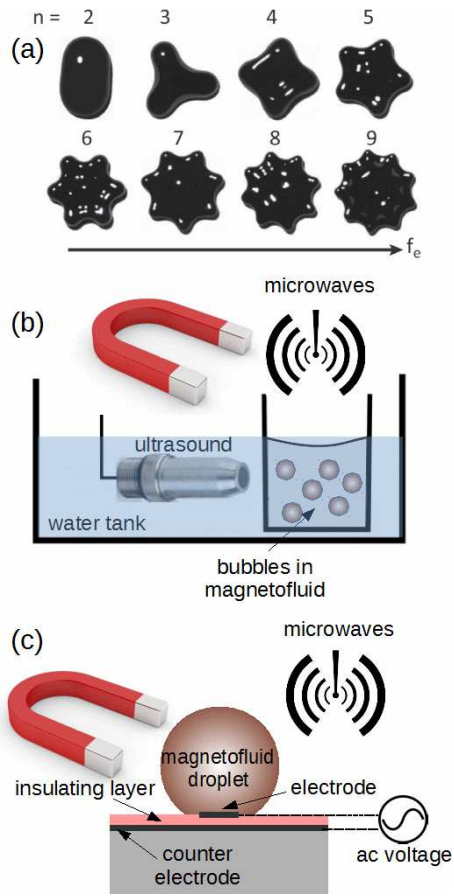}}
\caption{(a) Oscillation shapes of a magnetofluid droplet as a function of the forcing frequency $f_{\rm{e}}$. The number $n$ denotes the number of oscillating lobes (from left to right and top to bottom). Reproduced with permission from Ref.~\onlinecite{Jam16}. Schematic illustration of experimental setups for the excitation of shape oscillations of (b) bubbles in a magnetic fluid with ultrasound and (c) magnetofluid droplets with ac voltage. In (b), an immersible ultrasonic transducer is placed into a water tank and a cuvette with bubbles (or droplets) is placed in the area of the focal spot of the transducer. The scheme in (c) implements the scenario of electrowetting, i.e. the electrical control of wettability used to handle liquid droplets \cite{Oh08}.}
\label{fig:fig6}
\end{figure}

This approach may be adopted at microwave frequencies by using magnetic fluids. A magnetic fluid is a colloidal suspension of ultrafine ferro- or ferrimagnetic nanoparticles suspended in a carrier liquid \cite{Jam16, Erd17}. Such a fluid is an electric insulator and it becomes strongly magnetised in the presence of a magnetic field. Thus, a solid YIG sphere can be substituted with a magnetic fluid droplet. Whereas the droplet will support the ferromagnetic resonance mode at typical GHz-range frequencies \cite{Fan04}, thereby creating conditions for a strong coupling between magnons and GHz-range microwave photons, dramatic oscillations of its shape [Fig.~\ref{fig:fig6}(a)] due to sound pressure [Fig.~\ref{fig:fig6}(b)] or electrical force [Fig.~\ref{fig:fig6}(c)] applied to the droplet \cite{Kva17} would result in a triple-resonance photon-magnon-phonon interaction similar to that previously demonstrated in a solid-state YIG system \cite{Zha16_1}.

It is noteworthy that intriguing magneto-optical effects have been observed in magnetic fluids \cite{Shu17}. This opens up opportunities for magneto-optical coupling between light and hybridised photon-magnon-phonon states. Microwave magnetic field applied to a magnetic fluid also leads an increase in the temperature of the droplet (or liquid surrounding the bubble), which in turn affects the oscillation frequency of the droplet (bubble) by modifying the liquid viscosity \cite{Ros02}. Thus, the use of magnetic fluids fits very well into the conceptual picture of Fig.~\ref{fig:fig0}.

A rigorous analysis of multiresonant magnetofluidic systems requires complex numerical methods \cite{Erd17}. However, in a first approximation, the sound-driven oscillations of air bubbles in a magnetic fluid can be analysed by considering a simple Rayleigh-Plesset equation \cite{Mak16_2}. A similar strategy may be adopted to analyse liquid droplets \cite{Jam16, Mak17}. By using this approach, one can estimate that a $0.5$-$\mu$m-diameter magnetofluid droplet tuned on its fundamental mode should oscillate at the same acoustic frequency as a $250$-$\mu$m-diameter YIG sphere \cite{Zha16_1}. Higher oscillation frequencies can be achieved with the same droplet tuned on one of its higher-order modes. The resonance frequency $f_{\rm{0}}$ of a gas bubble of radius $R_{\rm{0}}$ in a magnetofluid is approximately given by the formula $f_{\rm{0}}R_{\rm{0}} = 3$~m/s \cite{Mak16_2}. Thus, a $0.5$-$\mu$m-diameter bubble should also oscillate at around the same acoustic frequency as a $250$-$\mu$m-diameter YIG sphere \cite{Zha16_1}. In these estimations, the material parameters of the magnetic fluid were taken from Refs.~\onlinecite{Kor08, Kva17}.

\section{Conclusions}

In recent years, many of the new phenomena in the field of strong microwave photon-magnon coupling have been driven by systems based on metal microwave resonators loaded with a magnetic material. In this perspective, I have demonstrated that dielectric antennas made of or containing magneto-insulating materials can operate as strongly-coupled system. As compared with metal-based structures, such systems are characterised by low dissipation losses and also enable a strong local microwave magnetic field enhancement, thereby additionally increasing the photon-magnon coupling strength. I have shown that dielectric antennas also open up exciting and practically important opportunities to control the photo-magnon coupling with light, vibrations and sound, heat, and electric currents and fields. I have also described several perspective research directions where the strong photon-magnon coupling extends the capability of existing devices or provide an entirely new functionality.

Finally, it is worth noting that the actual application range of dielectric antennas extends well beyond the discussion presented in this perspective. For example, dielectric antennas have been proposed to be used \cite{Ioa17, McA18} in high-sensitivity haloscopes that aim to detect axions -- hypothetical elementary particles and attractive dark matter candidates -- via their coupling to photons \cite{Car13}. 

\begin{acknowledgments}
This work was supported by the Australian Research Council (ARC) through its Centre of Excellence for Nanoscale BioPhotonics (CE140100003). The author would like to thank Andy Greentree (RMIT University), Mikhail Kostylev (University of Western Australia), Sergei Turitsyn (Aston University), Gennadiy Churyumov (NURE University), Igor Magda (Kharkov Institute of Physics and Technology) for valuable discussions. 
\end{acknowledgments}

\bibliography{strong_coupling_arxiv.bbl}

\end{document}